\documentclass{ws-procs975x65}
\def\be{\begin{equation}}
\def\ee{\end{equation}}

\newcommand{\keV}{\mbox{keV}}
\newcommand{\GeV}{\mbox{GeV}}

\begin{document}

\title{DARK MATTER: THE CASE OF STERILE NEUTRINO\footnote{
Based on talks given at 11th Marcel Grossmann Meeting on General
Relativity  (Berlin, 23.7 - 29.7.2006), at XXXIII International
Conference on High Energy Physics (Moscow, 26.7-2.7.2006), and at 6th
International Workshop on the Identification of Dark Matter (Rhodes,
11.9-16.9.2006). To appear in the Proceedings of Marcel Grossmann Meeting.}}

\author{Mikhail Shaposhnikov}

\address{Institut de Th\'eorie des Ph\'enom\`enes Physiques,
Ecole Polytechnique F\'ed\'erale de Lausanne,
CH-1015 Lausanne, Switzerland}


\begin{abstract}
An extension of the Standard Model by three right-handed neutrinos
with masses smaller than the electroweak scale (the $\nu$MSM) can
explain simultaneously dark matter and baryon asymmetry of the
Universe, being consistent with the data on neutrino oscillations. A
dark matter candidate in this theory is the sterile neutrino with the
mass in keV range. We discuss the constraints on the properties of
this particle and mechanisms of their cosmological production. Baryon
asymmetry generation in this model is reviewed. Crucial experiments
that can confirm or rule out the $\nu$MSM  are briefly discussed.
\end{abstract}

\bodymatter

\section{Introduction}
There is compelling evidence that the Minimal Standard Model (MSM) of
strong and electroweak interactions is not complete. There are
several {\em experimental} facts that cannot be explained by the MSM.
These are neutrino oscillations, the presence of dark matter in the
Universe, the baryon asymmetry of the Universe, its flatness, and the
existence of cosmological perturbations necessary for structure
formation. Indeed, in the MSM neutrinos are strictly massless and do
not oscillate. The MSM does not have any candidate for non-baryonic
dark matter. Moreover, with the present experimental limit on the
Higgs mass, the high-temperature phase transition, required for
electroweak baryogenesis, is absent. In addition, it is a challenge
to use CP-violation in  Kobayashi-Maskawa mixing of quarks to produce
baryon asymmetry in the MSM. Finally, the couplings of the single
scalar field of the MSM are too large for the Higgs boson to play the
role of the inflaton. This means that the MSM is unlikely to be a
good effective field theory up to the Planck scale.

In\cite{Asaka:2005an,Asaka:2005pn,Shaposhnikov:2006xi} it was
proposed  that a simple extension of the MSM by three singlet
right-handed neutrinos and by a real scalar field (inflaton) with
masses smaller than the electroweak scale may happen to be  a correct
effective theory up to some high-energy scale, which may be as large
as the Planck scale.  This model was called ``the $\nu$MSM",
underlying the fact that it is the extension of the MSM in the
neutrino sector. Contrary to Grand Unified Theories, the $\nu$MSM
does not have any internal hierarchy problem, simply because it is a
theory with a single mass scale. Moreover, as the energy behaviour of
the gauge couplings in this theory is the same as in the MSM, the
absence of gauge-coupling unification in it indicates that there may
be no grand unification, in accordance with our assumption of the
validity of this theory up to the Planck scale. As well as the MSM,
the $\nu$MSM does not provide any explanation why the weak scale is
much smaller than the Planck scale. Similarly to the MSM, all the
parameters of the $\nu$MSM can be determined experimentally since
only accessible energy scales are present.

As we demonstrated in\cite{Asaka:2005an,Asaka:2005pn}, the $\nu$MSM 
can explain simultaneously dark matter and baryon asymmetry of the
Universe being consistent with neutrino masses and mixings observed
experimentally. Moreover, in\cite{Shaposhnikov:2006xi} we have shown
that inclusion of an inflaton with scale-invariant couplings to the
fields of the $\nu$MSM allows us to have inflation and provides a
common source for electroweak symmetry breaking and Majorana neutrino
masses of singlet fermions -- sterile neutrinos. The role of the dark
matter is played by the lightest sterile neutrino with mass $m_s$ in
the  keV range. In addition,  the coherent oscillations of two other,
almost degenerate, sterile neutrinos  lead to the creation of baryon
asymmetry of the Universe\cite{Asaka:2005pn}  through the splitting
of the lepton number between active and sterile
neutrinos\cite{Akhmedov:1998qx} and electroweak
sphalerons\cite{Kuzmin:1985mm}. For review of other astrophysical
applications of sterile neutrinos see talk by Peter Biermann at this
conference\cite{Biermann:2007ap}.

In this talk I review the structure of the $\nu$MSM and discuss its
dark matter candidate -- sterile neutrino. The baryogenesis in this
model is briefly reviewed.

\section{The $\nu$MSM}
If three singlet right-handed fermions $N_I$ are added to the
Standard Model, the most general renormalizable Lagrangian describing
all possible interactions has the form:
\be
L_{\nu MSM}=L_{MSM}+
\bar N_I i \partial_\mu \gamma^\mu N_I
  - F_{\alpha I} \,  \bar L_\alpha N_I \epsilon \Phi^*
  - \frac{M_I}{2} \; \bar {N_I^c} N_I + h.c.,
  \label{lagr}
  \ee
where $L_{MSM}$ is the Lagrangian of the MSM, $\Phi$ and $L_\alpha$
($\alpha=e,\mu,\tau$) are the Higgs and lepton doublets,
respectively, and both Dirac ($M^D = f^\nu \langle \Phi \rangle$) and
Majorana ($M_I$) masses for neutrinos are introduced. In comparison
with the MSM, the $\nu$MSM contains 18 new parameters: 3 Majorana
masses of new neutral fermions $N_i$,  and 15 new Yukawa couplings in
the leptonic sector (corresponding to 3 Dirac neutrino masses, 6
mixing angles and 6 CP-violating phases).

Let us discuss in general terms what kind of scale for Majorana
neutrino masses $M_I$ one could expect. If Dirac neutrino masses
$(M^D)_{\alpha I}= F_{\alpha I} v$ (where $v=174$ GeV is the vacuum
expectation value of the Higgs doublet) are much smaller than the
Majorana masses $M_I$, the see-saw formula for active neutrino
masses 
\be
m_\nu = - M^D\frac{1}{M_I}[M^D]^T~
\label{eq:Mseesaw}
\ee
is valid.  Though it is known that the masses of active neutrinos are
smaller than $O(1)$ eV, it is clear that the scale of Majorana
neutrino masses cannot be extracted. This is simply because  the
total number of physical parameters describing  $m_\nu$ is equal to
$9$ (three absolute values of neutrino masses, three mixing angles
and three CP-violating phases), which is two times smaller than the
number of new parameters in the $\nu$MSM.

A most popular proposal\cite{Seesaw} is to say that the Yukawa
couplings $F$ in the active-sterile interactions are of the same
order of magnitude as those in the quark and charged lepton sector. 
This choice is usually substantiated by aesthetic considerations, but
is not following from any experiment. Then one has to introduce a new
energy scale,  $M_I \sim 10^{10}-10^{15}$ GeV, which may be related
to grand unification. The model with this choice of $M_I$ has several
advantages in comparison with the MSM: it can explain neutrino masses
and oscillations, and give rise to baryon asymmetry of the Universe
through leptogenesis\cite{Fukugita:1986hr} and anomalous electroweak
number non-conservation at high temperatures\cite{Kuzmin:1985mm}.
However, it cannot explain the dark matter as the low energy limit of
this theory is simply the MSM with non-zero active neutrino masses
coming from dimension five operators. On a theoretical side, as a
model with two very distinct energy scales it suffers from a
fine-tuning hierarchy problem $M_I \gg M_W$. Also, since the energy
scale which appears in this scenario is so high, it would be
impossible to make a direct check of this conjecture by experimental
means.

Another suggestion is to fix the Majorana masses of sterile neutrinos
in $1-10$ eV energy scale\cite{deGouvea:2005er} to accommodate the
LSND anomaly\cite{Aguilar:2001ty}. The theory with this choice of
parameters, however, cannot explain the baryon asymmetry of the
Universe and does not provide a candidate for dark matter particle.

Yet another paradigm is to determine the parameters of the $\nu$MSM
from available observations, i.e. from  requirement that it should
explain neutrino oscillations, dark matter and baryon asymmetry of
the universe in a unified way. It is this choice that will be
discussed below. It does not require introduction of any new energy
scale, and  $M_I < M_W$. In this case the Yukawa couplings must be
much smaller than those in the quark sector, $F < 10^{-6}$. The
theory has a number of directly testable predictions, which can
confirm or reject it.

\section{Dark matter}
Though the $\nu$MSM  does not offer any stable particle besides those
already present in the MSM, it contains a sterile neutrino with a
life-time exceeding the age of the Universe, provided the
corresponding Yukawa coupling is small enough. The decay rate of
$N_1$ to three active neutrinos and antineutrinos (assuming that
$N_1$ is the lightest sterile neutrino) is given by 
\be
 \Gamma_{3 \nu}
  = \frac{G_F^2 \, M_1^5 \theta^2}{96 \, \pi^3}~,~~
  \theta=\frac{m_0}{M_1}~,
  ~~m_0^2=\sum_{\alpha = e, \mu, \tau}
  |M^D{\alpha 1}|^2~,
   \label{eq:G_3nu}  
\ee
where $G_F$ is the Fermi constant. For example, a choice of $m_0 \sim
O(1)$ eV and of $M_1 \sim O(1)$ keV leads to a sterile neutrino
life-time $\sim 10^{17}$ years\cite{Dolgov:2000ew}.

The mass of the sterile dark matter neutrino cannot be too small. An
application of the Tremaine-Gunn arguments\cite{Tremaine:1979we} to
the dwarf spheroidal galaxies\cite{Lin:1983vq} gives the lower
bound\cite{Dalcanton:2000hn} $M_1 > 0.3~\mbox{keV}$. 
If the sterile neutrino mass is in the keV region,
it may play a role of warm dark
matter\cite{Dodelson:1993je,Abazajian:2001nj}. Sterile neutrino free
streaming length an matter-radiation equality is given by
\be
\lambda_{FS} \simeq 1~\mbox{Mps}\left(\frac{1~\keV}{M_1}\right)
\left(\frac{\langle p_s/T \rangle}{3.15}\right)
\ee
and the mass inside $\lambda_{FS}$ is
\[
M_{FS} \simeq 3\times 10^{10} M_\odot
\left(\frac{1~\keV}{M_1}\right)^3
\left(\frac{\langle p_s/T \rangle}{3.15}\right)^3~,
\]
where  $<p_s>~(<p_a>)$ is an average momentum of sterile (active)
neutrino at the moment of structure formation, $M_\odot$ is the solar
mass. One normally defines cold dark matter (CDM) as that
corresponding to  $M_{FS} < 10^{5}M_\odot$, hot DM as the one with 
$M_{FS} > 10^{14}M_\odot$, and warm DM as anything in between.
Potentially, WDM could solve some problems of the CDM scenario, such
as the missing satellites problem\cite{Moore:1999nt,Bode:2000gq} and
the problem of cuspy profiles in the CDM
distributions\cite{Goerdt:2006rw,Gilmore:2006iy}.

Even stronger constraint  on the mass of sterile neutrino comes from
the analysis of the cosmic microwave background and the matter power
spectrum inferred from Lyman-$\alpha$ forest
data\cite{Hansen:2001zv,Viel:2005qj}
: $M_1 >  M_0\left(\frac{<p_s>}{<p_a>}\right)$.  According
to\cite{Seljak:2006qw}, $M_0 = 14.5$ keV, whereas\cite{Viel:2006kd}
gives $M_0 = 10$ keV.

Yet another constraint on the parameters of dark matter sterile
neutrino comes from radiative decay $N_1 \to \nu\gamma$, suppressed
in comparison with $N_1 \to 3\nu$ by a factor $O(\alpha)$ ($\alpha$
is a fine structure constant). This two body decay produces a line in
the spectrum of X-rays coming from dark matter in the Universe;
corresponding constraints are discussed in detail in other
contributions to these proceedings, see also Refs.
\cite{Abazajian:2001vt,Boyarsky:2005us,Boyarsky:2006zi,
Boyarsky:2006fg,Riemer-Sorensen:2006fh,Watson:2006qb,Riemer-Sorensen:2006pi,
Boyarsky:2006ag,Boyarsky:2006kc,Abazajian:2006jc,Boyarsky:2006hr}. To get an idea on
admitted Yukawa coupling constants and of Dirac mass of dark matter
sterile neutrino see Fig. 1, based on results of Ref.
\cite{Boyarsky:2006fg,Boyarsky:2006ag}
\begin{figure}[t]
\epsfysize=9cm
\centerline{\epsffile{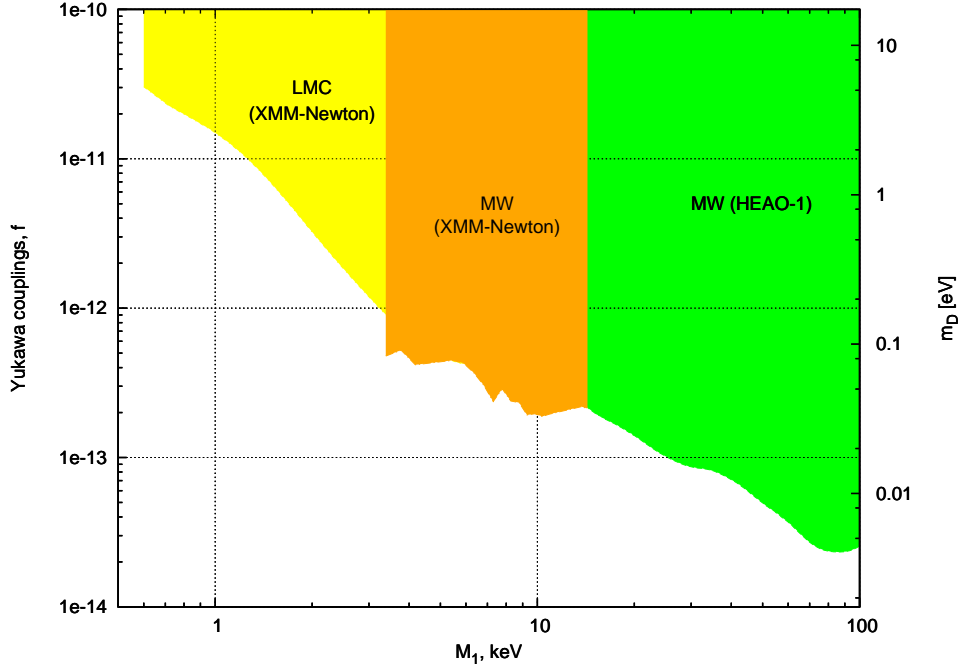}}
\caption{Upper bound on Yukawa coupling constant (left vertical axis)
and Dirac mass (right vertical axis) of dark matter sterile neutrino,
coming from X-ray observations of Large Magellanic Cloud (LMC) and
Milky Way (MW) by XMM-Newton and HEAO-1 satellites.}
\end{figure}

\section{Cosmological production of sterile neutrinos}
Let us discuss now cosmological production of sterile neutrinos. In
the region of the parameter space admitted by X-ray observations
sterile neutrinos were never in thermal equilibrium in the early
Universe\cite{Dodelson:1993je}. This means that their abundance
cannot be predicted in the framework of the $\nu$MSM
\cite{Asaka:2006rw}: one should either fix the concentration of
sterile neutrinos at temperatures greater than $1$ GeV, or specify
the physics beyond the $\nu$MSM.

One can address the question how many sterile neutrinos are produced
due to the $\nu$MSM interactions, eq. (\ref{lagr}), i.e. because of
the mixing with active neutrino flavours characterized by parameter
$\theta$. In fact, this mixing is temperature dependent
\cite{Barbieri:1989ti}:
\[
\theta \rightarrow \theta_M \simeq
\frac{\theta}{1+2.4(T/200~\rm{MeV})^6(\rm{keV}/M_1)^2}~,
\]
so that the rate $\Gamma$ of sterile neutrino production  is strong
suppressed at $T> 100$ MeV, $\Gamma \propto T^{-7}$. The rate peaks
roughly at \cite{Dodelson:1993je} $T_{peak} \sim
130\left(\frac{M_I}{1~\keV}\right)^{1/3}~\mbox{MeV}$, which
corresponds to the temperature of the QCD cross-over for keV scale
sterile neutrinos. This fact makes an exact estimate of the number of
produced sterile neutrinos to be a very difficult task
(see\cite{Asaka:2006rw} for a discussion of the general formalism for
computation of sterile neutrino abundance), since $T_{peak}$ happens
to be exactly at the point where the quark-gluon plasma is strongly
coupled and the dilute hadron gas picture is not valid. The chiral
perturbation theory works only at  $T < 50$ MeV. The perturbation
theory in QCD works only at $T\gg \Lambda_{QCD}$, and the convergence
is very slow. The lattice simulations work very well for pure
gluodynamics. However, no results with three light quarks and with
reliable extrapolation to continuum limit are available yet. Also,
the treatment of hadronic initial and final states in reactions $\nu
+ q \to \nu + q ,~ q + \bar{q} \to \nu \bar{\nu}$ is quite uncertain.
In refs. \cite{Dodelson:1993je,Dolgov:2000ew} the computation of
sterile neutrino production was done with the use of simplified
kinetic equations and without accounting for hadronic degrees of
freedom. In \cite{Abazajian:2001nj,Abazajian:2005xn} some effects
related to existence of quarks and hadrons in the media were
included; the same type of kinetic equations were used. In\cite{Asaka:2006nq}
a computation of sterile neutrino production based on first
principles of statistical physics and quantum field theory has been
done and uncertainties related to hadronic dynamics were analyzed.
The results are presented in Fig. 2. They correspond to the case when
there is no entropy production ($S=1$) due to decay of heavier
sterile neutrinos of the $\nu$MSM\cite{Asaka:2006ek}. The area above dotted line is
certainly excluded: the amount of produced dark matter would lead to
over-closer of the universe. The region below dashed line is
certainly allowed: the amount of sterile neutrinos produced due to
active-sterile transitions is smaller than the amount of dark matter
observed. Any point in the region between two solid lines
(corresponding to the ``most reasonable" model for hadronic
contribution\cite{Asaka:2006nq}) can lead to dark matter generation entirely
due to active-sterile transitions.  Maximal variation of the hadronic
model, defined in\cite{Asaka:2006nq} extends this region to the space between 
dotted and dashed lines. In the case of entropy production with $S>1$
all these four lines simply move up by a factor $S$.
\begin{figure}[t]
\epsfysize=10cm
\centerline{\epsffile{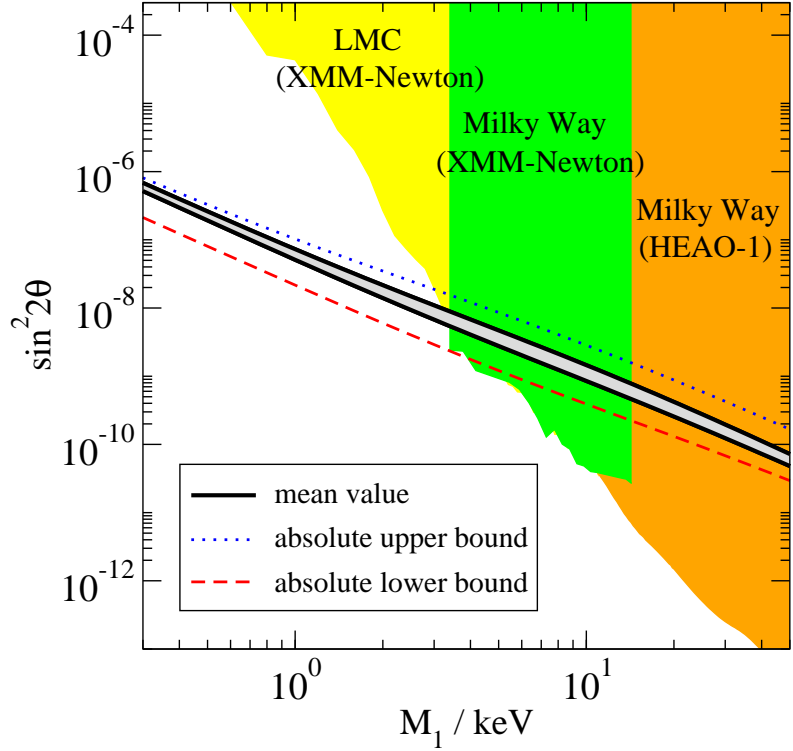}}
\caption{X-ray constraints from
\cite{Boyarsky:2006fg,Boyarsky:2006ag} versus required mixing of
sterile neutrino in Dodelson-Widrow scenario. It is assumed that no
entropy production from decays of heavier sterile neutrinos of the
$\nu$MSM is taking place. The area between two solid lines
corresponds to all possible variations of mixing angles to different
leptonic families for ``best choice" hadronic dynamics\cite{Asaka:2006nq}. The
area between dotted and dashed lines corresponds to most conservative
estimate of hadronic uncertainties\cite{Asaka:2006nq}.}  
\end{figure}

One can see that the active-sterile mixing can accommodate for all
dark matter only if $M_1 < 3.5$ keV, if the ``most reasonable"
hadronic model is taken. The most conservative limit would correspond
to  $M_1 < 6$ keV, if all hadronic uncertainties are pushed in the
same direction and the uncertainty by a factor of $2$ is admitted for
the X-ray bounds. Therefore, if Lyman-$\alpha$ constraints
of\cite{Seljak:2006qw,Viel:2006kd} are taken for granted, the
production of sterile neutrinos due to active-sterile neutrino
transitions happens to be too small to account for observed abundance
of dark matter. In other words, physics beyond the $\nu$MSM is likely
to be required to produce dark matter sterile neutrinos. Another
option is to assume that the universe contained relatively large
lepton asymmetries\cite{Shi:1998km}.

In\cite{Shaposhnikov:2006xi} it was proposed the the $\nu$MSM may be
extended by a light inflaton in order to accommodate inflation. To
reduce the number of parameters and to have a common source for the
Higgs and sterile neutrino masses the  inflaton-$\nu$MSM couplings
can be taken to be scale invariant on the classical level:
\be
  {\cal L}_{\nu \rm{MSM}}\rightarrow
   {\cal L}_{\nu \rm{MSM}[M\rightarrow 0]} + 
  \frac{1}{2}(\partial_\mu\chi)^2 -
  \frac{f_I}{2} \; \bar {N_I}^c  N_I \chi + \rm{h.c.} - V(\Phi,\chi)~,
\ee
where the Higgs-inflaton potential is given by:
\[
V(\Phi,\chi) =
\lambda\left(\Phi^\dagger\Phi-\frac{\alpha}{\lambda}\chi^2\right)^2+
\frac{\beta}{4}\chi^4 -\frac{1}{2} m_\chi^2 \chi^2 .
\]
The requirement that the chaotic inflation\cite{Linde:1983gd} produces the
correct amplitude for scalar perturbations leads to the constraints:
\[
\beta \simeq  10^{-13}, ~~~ \alpha \lesssim 10^{-7},~~~f_I \lesssim
10^{-3}~.
\]
For $\alpha > \beta$ inflaton mass is smaller than the Higgs mass,
$m_I < M_H$.

One can show\cite{Shaposhnikov:2006xi} that the  
inflaton with mass $m_I > 300$ MeV is in thermal equilibrium  thanks
to reactions $\chi \leftrightarrow e^+ e^-,~ \chi \leftrightarrow
\mu^+ \mu^-$ down to $T < m_I$. The 
sterile neutrino abundance due to inflaton decays: $\chi \to NN$ is
given by
\[
\Omega_s \simeq 0.26
 \frac{\Gamma M_0 m_s}{m_I^2 \times 12~\rm{eV}}\frac{2\pi
 \zeta(5)}{\zeta(3)}~.
\]
So, for $m_I \sim 300$ MeV ($m_I \sim 100$ GeV) the correct
$\Omega_s$ is obtained for $m_s \sim 16-20$ keV ($m_s \sim {\cal O}
(10)$ MeV).
A sterile neutrino in this mass range is perfectly consistent with
all cosmological and astrophysical observations. As for the bounds on
mass versus active--sterile mixing coming from X-ray observations of
our galaxy and its dwarf satellites, they are easily
satisfied since the production mechanism of sterile neutrinos
discussed above has nothing to do with the active--sterile neutrino
mixing leading to the radiative mode of sterile neutrino decay.

\section{Baryon Asymmetry of the Universe}
The  baryon (B) and lepton
(L) numbers are not conserved in the $\nu$MSM. The lepton number is
violated by the Majorana neutrino masses, while  $B+L$ is broken by
the electroweak anomaly. As a result, the sphaleron processes with
baryon number non-conservation are in thermal equilibrium  for 
$100$ GeV $< T < 10^{12}$ GeV. As for CP-breaking, the $\nu$MSM
contains  $6$ CP-violating phases in the lepton sector and a
Kobayashi-Maskawa phase in the quark sector. This makes two of the
Sakharov conditions\cite{Sakharov:1967dj} for baryogenesis satisfied.
Similarly to the MSM, this theory does not have an electroweak phase
transition with allowed values for the Higgs
mass\cite{Kajantie:1996mn}, making impossible the electroweak
baryogenesis, associated with the non-equilibrium bubble expansion.
However, the $\nu$MSM contains extra degrees of freedom - sterile
neutrinos - which may be out of thermal equilibrium exactly because
their Yukawa couplings to ordinary fermions are very small. The
latter fact is a key point for the baryogenesis in the $\nu$MSM,
ensuring the validity of the third Sakharov condition. 

In\cite{Akhmedov:1998qx} it was proposed that the baryon asymmetry
can be generated through CP-violating sterile neutrino oscillations.
For small Majorana masses the total lepton number of the system,
defined as the lepton number of active neutrinos plus the total
helicity of sterile neutrinos, is conserved and equal to zero during
the Universe's evolution. However, because of oscillations the lepton
number of active neutrinos becomes different from zero and gets
transferred to the baryon number due to rapid sphaleron transitions.
Roughly speaking, the resulting baryon asymmetry is equal to the
lepton asymmetry at the sphaleron freeze-out. 

The kinetics of sterile neutrino oscillations and of the transfers
of  leptonic number between active and sterile neutrino sectors has
been worked out in\cite{Asaka:2005pn}.  The effects to be taken into
account include oscillations, creation and destruction of sterile and
active neutrinos, coherence in sterile neutrino sector and its lost
due to interaction with the medium, dynamical asymmetries in active
neutrinos and charged leptons. The corresponding equations are
written in terms of the density matrix for sterile neutrinos and
concentrations of active neutrinos and are rather lengthy and will
not be presented here due to the lack of space. They can be found in
the original work\cite{Asaka:2005pn}. The corresponding equations are
to be solved with the choice of the $\nu$MSM parameters consistent
with the experiments on neutrino oscillations and with the
requirement that dark matter neutrino has the necessary properties.

The value of baryon to entropy ratio $\frac{n_B}{s}$ can be found
from the solution of the kinetic equations and is given
by\cite{Asaka:2005pn}
\begin{eqnarray}
  \frac{n_B}{s} \simeq
    1.7\cdot 10^{-10} \, {\delta_{\rm CP} }\,
 {\left( \frac{10^{-5}}{\Delta M_{32}^2/M^2_3}
 \right)^{\frac{2}{3}}}
 {\left( \frac{M_3}{10~ \GeV } \right)^{\frac{5}{3}}} \,,
 \nonumber
\end{eqnarray}
where $M_{2,3}$ are the masses of the heavier sterile neutrinos,
$\Delta M_{32}^2 = M_3^2-M_2^2$, and the CP-breaking factor
$\delta_{\rm CP}$ is expressed through the different mixing angles
and CP-violating phases, parameterizing the Dirac neutrino masses, and
can be ${\cal O}(1)$, given the present experimental data on neutrino
oscillations. This shows that the correct baryon asymmetry of the
Universe $\frac{n_B}{s} \simeq (8.8-9.8)\times 10^{-11}$ is generated
when the heavier sterile neutrinos with the masses, say, $1$ GeV are
degenerate to one part in $10^5$. This looks like a strong fine
tuning but may also indicate that there exists some symmetry making
the degeneracy automatic\cite{Shaposhnikov:2006nn}.

It is interesting to note that for masses of sterile neutrinos $>
100$ GeV the mechanism does not work as the sterile neutrinos
equilibrate. Also, the temperature of baryogenesis is rather low,
$T_L \simeq (\Delta M^2 M_{Pl})^{\frac{1}{3}}> 10^{2}$ GeV, i.e.
validity of $\nu$MSM is only required at the scales smaller than
$M_W$ or so.

\section{Conclusions}
 The $\nu$MSM is, perhaps, the simplest and the most economical
extension of the Minimal Standard Model. It shares with the MSM its
advantages (renormalizability and agreement with most particle
physics experiments) and its fine-tuning problems (the gauge
hierarchy problem, flavour problem, etc). However, unlike the MSM,
the $\nu$MSM  can explain simultaneously three different  phenomena,
observed experimentally, namely neutrino oscillations, dark matter,
and baryon asymmetry of the Universe.  The parameter-space of the
model is rather constrained: the dark matter neutrino should have a
mass in the keV region and be much lighter than two heavier sterile
neutrinos, which are required to be quite degenerate. The model has a
number of testable predictions. In astrophysics, one should search
for  X-rays from decays of dark matter neutrinos, which could be
achieved with a X-ray spectrometer in Space with good energy
resolution $\delta E/E \sim 10^{-3}-10^{-4}$ getting signals from our
Galaxy and its Dwarf
satellites\cite{Boyarsky:2006fg,Boyarsky:2006hr}. In particle
physics, the $\nu$MSM predicts the absolute values of active neutrino
masses\cite{Asaka:2005an,Boyarsky:2006jm}, and existence of
relatively light singlet fermions\cite{Shaposhnikov:2006nn} which can
be searched for in decays of charmed, beauty and even $K$ or
$\pi$-mesons in experiments similar
to\cite{Bernardi:1987ek,Astier:2001ck}. The dark matter neutrino can
be looked for in $\beta$-decays of tritium and other
isotopes\cite{Bezrukov:2006cy}.

\section*{Acknowledgments}
This work was supported in part by the Swiss National Science
Foundation. It is a pleasure to thank Takehiko Asaka, Fedor Bezrukov,
Steve Blanchet, Alexey Boyarsky, Alexander Kusenko, Mikko Laine,
Andrei Neronov, Oleg Ruchayskiy, and Igor Tkachev for collaboration.



\end{document}